\def\eps@scaling{0.9}
\def\plotone#1{\centering \leavemode
\epsfxsize=\eps@scaling\textwidth \epsfbox{#1}}
\def\lsim{\lower 4pt\vbox{\hbox{$\buildrel < \over \sim$ }}}
\def\gsim{\lower 4pt\vbox{\hbox{$\buildrel > \over \sim$ }}}
\def\eqnumber{\eqno(\the\count1)\global\advance\count1 by 1}
\input{epsf.sty}
\documentstyle[11pt,aaspp4]{article}

\slugcomment{To appear in The Astronomical Journal}

\lefthead{Reed et al.}
\righthead{Expansion of NGC 6543}

\begin{document}

\title{HST Measurements of the Expansion of NGC~6543: Parallax
Distance and Nebular Evolution
\footnote{Based on observations made
with the NASA/ESA Hubble Space Telescope, obtained from the data
archive at the Space Telescope Science Institute.  STScI is operated
by the Association of Universities for Research in Astronomy, Inc.\
under NASA contract NAS 5-26555.}}

\author{Darren S.\ Reed \& Bruce Balick}
\affil{Astronomy Department, University of Washington,
Seattle, WA}

\author{Arsen R.\ Hajian \& Tracy L.\ Klayton} 
\affil{Department of Astrometry, United States Naval Observatory,
Washington, D.C.}

\author{Stefano Giovanardi}  
\affil{Universit\`{a} degli Studi di Bologna}

\author{Stefano Casertano \& Nino Panagia} 
\affil{Space Telescope Science Institute,}
\affil{On assignment from the Space Science Department of ESA.}

\and

\author{Yervant Terzian}
\affil{Department of Astronomy and NAIC, Cornell University, Ithaca, NY}

\begin{abstract}

The optical expansion parallax of NGC~6543 has been detected and
measured using two epochs of HST images separated by a time baseline
of only three years. We have utilized three separate methods of
deriving the angular expansion of bright fiducials, the results of
which are in excellent agreement. We combine our angular expansion
estimates with spectroscopically obtained expansion velocities to
derive a distance to NGC~6543 of 1001$\pm$269 pc.  The deduced
kinematic age of the inner bright core of the nebula is 1039$\pm$259
years; however, the kinematic age of the polar caps that surround the
core is larger - perhaps the result of deceleration or earlier mass
ejection.  The morphology and expansion patterns of NGC~6543 provide
insight into a complex history of axisymmetric, interacting stellar mass
ejections.

\end{abstract}

\keywords{planetary nebulae: individual (NGC 6543) --- astrometry }

\section{Introduction}

Accurate distances to planetary nebulae (PNe) are critical for
calculating the size, mass, luminosity, age, and other properties of
PNe and their central stars.  Distances to PNe are also crucial when
studying properties of the galaxy such as the galactic rotation, and
scale height of their progenitor stars.  In addition, distances are
required to ascertain the planetary nebula luminosity function in
order to use PNe as standard candles for cosmological distance studies
(Ciardullo {\it et al.}\ 1988).

Despite being bright and containing (dominated by) a rich line
spectrum, PNe have distances that are remarkably ill-constrained.  In
fact, distances to many extragalactic PNe are better known than
distances to most galactic PNe.  Many different distance measurement
techniques have been applied to PNe in the past. Most of have been
statistical in nature, presuming that some property is common to
the entire population (Hajian \&
Terzian 1996). The most commonly used PN distance estimation technique
is known as the Shklovsky method (Shklovsky 1956), and assumes a
constant ionized gas mass for all PNe.  Unfortunately, these statistical
techniques yield errors that are large, often a factor of two or more, as
reviewed by Terzian (1997).  In the case of the Shklovsky method, $D
\propto M_{\rm i}^{0.4}$, where $D$ is the nebular distance and
$M_{\rm i}$ the ionized mass.  Since PNe progenitor masses can span a
factor of $\approx$10, and assuming that $M_{\rm i}$ scales with the
progenitor mass, distance errors of several hundred percent are
typical.

The most reliable distance measurements for PNe so far have been
provided by measuring their expansion parallaxes (Terzian 1997).  PN
shell expansion velocities are typically $\sim10$ km s$^{-1}$, or
$\sim2$ mas yr$^{-1}$ at distances of $\sim1$ kpc (mas =
milliarcsecond).  Expansion parallax distance determinations require
two epochs of image data separated by a time baseline $\ga1$ yr, which
can be used to measure the angular expansion ``parallax'',
$\dot\theta$, of a fiducial feature. The physical expansion velocity
of the features must be spectroscopically obtained, and converted to a
tangential velocity, $v_{\perp}$ ({\it i.e.}, the component of the
velocity normal to the line of sight) using a spatiokinematic model of
the nebula. The distance to a PN can then be computed directly by
dividing $v_{\perp}$ by $\dot\theta$ (Hajian, Terzian, \& Bignell
1993).

Distances to nine PNe have been previously measured using radio images
from the Very Large Array (VLA) at different epochs ( Masson 1989a; Masson
1989b; Gomez {\it et al.}\ 1993; Hajian, Terzian, \& Bignell 1993; Hajian,
Terzian, \& Bignell 1995; Hajian \& Terzian 1996; Haryadi \& Seaquist
1998).  One past attempt has been made to measure the expansion of
NGC~6543 using the VLA.  However, this expansion parallax measurement was
unsuccessful due to insufficient image quality in the radio-wavelength
data (Hajian \& Terzian 1996).

The many well-defined, optically bright features of NGC~6543 serve as
ideal fiducials in measuring its angular expansion. The ability to
measure the expansions of individual features within this
extraordinary nebula provides us with a way to sort through the
nebula's recent history, to probe the mass ejection behavior of its
nucleus, and to directly determine the distance to this PN.

As discussed by Hajian, Terzian, \& Bignell (1995), the optical
expansion of a PN is particularly detectable and measurable with high
accuracy ($<$25\%) using the Hubble Space Telescope (HST).  In this
paper, we report the detection of the systematic expansion of NGC~6543
between 1994 and 1997 from HST archival observations.  In order to
quantify the magnitude of the angular expansion rate, we present three
separate astrometric analyses of these recorded images.  These
observations, along with a spatiokinematic model of the nebula derived
from independent measurements of the Doppler
velocity, permit three
direct distance determinations, all of which are in excellent mutual
agreement.

In addition, the expansion pattern reveals that the kinematic ages of
features approximately scale with their angular size.  This provides
insight into the highly organized and orchestrated mass loss history of
the nucleus, presently a WR-Of star (Heap \& Augensen 1987).  This
two-in-one paper discusses both the distance to and the nebular evolution
of NGC~6543.

\section{Observations}

The bright core of NGC~6543 just fills the higher-resolution planetary
camera (``PC'') of the Wide Field and Planetary Camera 2 (``WFPC2'')
whose projected pixel size is 45.5 mas. Archived unsaturated HST
narrowband images from GO program 5403 and corresponding calibration
images from GO program 6943 were used for this study (Table~1). The
data epochs are separated by 35 months.  Images through F502N (hereafter
``[O~III]''), F656N (``H$\alpha$''), and F673N (``[S~II]'') filters were
obtained at both epochs.  The use of the same instrumentation for the
two epochs, albeit at differing orientations, reduces systematic errors.

\placetable{tbl-1}

Measurements of the expansion were based solely on images in the
[O~III]$\lambda$502nm and [S~II]$\lambda$673nm lines.  The [O~III]
image is optimal for this study since it is characterized by
well-defined, bright structures with excellent signal-to-noise ratios
in the WFPC2 images.  The H$\alpha$ images appear essentially
redundant to their [O~III] counterparts.  For this reason the H$\alpha$
images were not analyzed. The [S~II] images were used to confirm the
expansion measurements from the [O~III] images, and to investigate the
proper motions of specific, bright, low-ionization features.

The pointing and orientation of the HST were different for the 1994
and 1997 observations. Therefore, after cosmic rays were removed and
the relative image intensities were calibrated, the images had to be
regridded to a common center and orientation.  We used the IRAF
tasks ``drizzle'', ``imshift'', and ``rotate'' to remove geometric
distortions from,
align, and rotate the images. The shifts and rotations were dithered
until patterns of residuals of the corresponding images from the two
epochs showed no trace of alignment or orientation errors. We estimate the
accuracy of the alignment to be $\la$ 0.05 pixel (2 mas) and the
orientation to be $\sim$ 0.1$^\circ$.

\section{Data Reduction And Results}
\subsection{Methodology}

Blinking of the 1994 and 1997 images shows clearly that the nebula has
changed its scale size ({\it i.e.} the pattern changes are radial).
Our strategy is therefore to average shifts in the locations of
fiducials on equal and opposite sides of the nucleus (along radial
lines) using various methods.

The angular expansion rate, $\dot\theta$, is best measured using
fiducials with narrow peaks and/or sharp edges.  In our analysis, we
concentrated on the edges of several nebular features shown in Figure 1.
Using the nomenclature of  Miranda \& Solf (1992; hereafter ``MS92''), we
analyzed the ``P.A. 25$^\circ$
ellipse'' (E25) and the ``P.A. 105$^\circ$ ellipse'' (E105).  In the
outer regions, we were able to measure the motion of the bright polar
caps and the faint polar condensations, referred to as (D-D$^\prime$)
and (F-F$^\prime$), respectively, in MS92. These structures are
identified in Figure 1.

We used three methods to extract $\dot\theta$ from the images. The
most direct approach, called the ``profile method'', is to compare
one-dimensional radial brightness distributions, or ``profiles'', from
images of two epochs taken with the same filter.  The relative
positions of sharp nebular features are found by analyzing the flux
gradients in the radial profiles in the residual image of the two epochs,
as pioneered by Masson (1986).  This method allows $\dot\theta$ to be
estimated along many one-dimensional ``cuts'' through the nebula.  The
random errors appear to be dominated by photon statistics.

A summary of the ``profile method'' can be found in Hajian,
Terzian, \& Bignell (1993, 1995) and in Hajian \& Terzian (1996),
referred therein as the ``expansion parallax algorithm''. The
analysis for HST images is identical to the analysis for VLA data
except that HST data are not obtained in the Fourier plane. The method
requires high signal-to-noise ratios in the residuals after the images
of two epochs are subtracted, and it assumes no change in surface
brightness or nebular shape.  Using this method, $\dot\theta$ can be
computed for any bright feature intersecting a radial cut:
\begin{equation}
\dot\theta _{profile} = {\Delta F\over \nabla F},
\end{equation}
where $\nabla F$ is the flux gradient of the nebular feature and
$\Delta F$ is the peak value of the difference map near the nebular
feature.  We take the uncertainty in expansion measurement along an
individual profile to be equal to (1) the difference between expansion
rates inferred from the corresponding features on opposite sides of the
nucleus, or (2) the unweighted mean and standard deviation of
multiple cuts intersecting an extended structure, where appropriate.

The ``radial fitting'' method is a similar and somewhat redundant
technique.  The positions of bright nebular features along 1-D radial
flux profiles are measured by performing Gaussian fits, by evaluating
the flux-weighted centroid of a fiducial's peak, or by graphically
measuring the shift of a sharp edge in each of the two epochs.

The radial fitting method has a somewhat increased potential for
yielding expansion estimates for nebular features that are too faint
for similar measurements with the profile method.  This is due to the
increased noise level in a difference map relative to the epoch maps
($\sqrt{2}$ for epoch maps with equal noise levels).  The random
errors of the measurement are dominated by the uncertainties in the
fit, which include the judgement involved in defining a fiducial.
For fits to peaks, the measurements of a Gaussian feature in a
complicated environment are sensitive to choices of the edges and base
of the fitted Gaussian curve. Great care is necessary to use the same
procedures for the images of both epochs.  Finding the relative positions
of sharp edges is straightforward.  It is done by overplotting the 1-D
flux profiles of the two epochs, and graphically measuring the radial
displacement at multiple positions along the steep flux dropoffs of the
edges.  We assume a minimum 20\% uncertainty in measured expansion rates
since we only employ the radial fitting method once for each fiducial
in each filter.

The ``magnification method'', is the simplest but least accurate of
the methods.  A magnification factor is applied to the first-epoch
image which minimizes the rms of the difference, or residual, image in
the regions of the bright fiducial features.

For each emission line, [O~III] and [S~II], we used the IRAF task
``geotran'' to magnify
the 1994 image by various magnification factors, $M$, in the range of
1.001 to 1.005.  For radial motions with no brightness changes, a
given feature will disappear in the residual image after the magnified
1994 image is subtracted from the aligned 1997 image.  We can then
compute $\dot\theta$ using the following equation:
\begin{equation}
\dot\theta _{mag,mas/yr} = {(M-1)\theta_{mas}\over 2.92~yr},
\end{equation}
where $\theta$ is the angular distance from the nucleus to the feature.

The advantages of the magnification method are that the entire image
is utilized, that the intuitive pattern recognition capability and
judgement of the brain can be exploited, and that the residuals may
uncover deviations from simple radial expansion in the complex nebula
that would not easily be recognized in one-dimensional analyses of the
brightness distribution.  The major disadvantage to the magnification
method is that errors are difficult to quantify.  We assume an uncertainty
of 25\% in our magnification method expansion rates.

\subsection{Distance Estimates and Kinematic Ages}

Once $\dot\theta$ has been measured using the above methods, and
$v_{\perp}$ has been determined based on spectroscopic observations
and a spatiokinematic model used to convert radial into tangential
velocities, the distance to the nebula in parsecs can be calculated:
\begin{equation}
D_{pc} = 211 \frac{v_{\perp,~km/s}}{\dot\theta_{mas/yr}}.
\end{equation}
The tangential component of the expansion velocities of various
features were taken from the detailed observations and the nebular
models of MS92.  In addition, the
kinematic age, $T$, of a fiducial is given by:
\begin{equation}
T_{yr} = \frac{\theta_{mas}}{\dot\theta_{mas/yr}}.
\end{equation}

We also note that a more accurate distance can be computed by taking all
of the measured values of $\dot\theta$ into account ({\it i.e.},
$\dot\theta$ at multiple P.A.s for multiple features, rather than just the
minor axis of E25, which our distance is based upon). However, since (1)
the only available kinematic information for NGC~6543 was obtained in an
emission line, [N~II]$\lambda6584$, that is different from the line(s)
used to determine the astrometric expansion of the nebula,
[O~III]$\lambda5007$ and [S~II]$\lambda6717+31$; and (2) uncertainty
in the adopted expansion velocity is approximately $\sim$10\%, we
did not feel it was necessary to fit a complex spatiokinematic model
to the data.  This will be more appropriate when we obtain
positionally resolved velocities in the summer of 1999.

\placetable{tbl-2}

We next discuss the measurements of $\dot\theta$ using all three
methods.  The results are compiled in Table~2.  Also shown in the table
is the expected value of $\dot\theta$ assuming uniform nebular expansion
whose rate is calibrated to the average measurement of $\dot\theta$ along
the minor axis of E25 (see below).

For the [O~III] images, we applied the profile analysis to various
radial cuts through the nucleus in $10^\circ$ increments, as shown in
Figure 2.  The resulting angular expansion rates are displayed in
units of mas yr$^{-1}$ wherever the profile intersects a bright, thin
feature.  For the [S~II] images we measured $\dot\theta$ along radial
cuts through the central star at intervals of 2 degrees as shown in
Figure 3.  The residuals of the magnification method are shown in
Figures 4 and 5, and the corresponding values of $\dot\theta$ are shown
in Table~2.

The changes in the nebular structure are seen most clearly in the
residual images of the unmagnified 1994-1997 image pairs.  See the upper
left panels of Figures 4 and 5. Although the residuals are only about 1\%
of the image brightnesses, the patterns of change are quite clear.
None of the residuals from any of these techniques yield evidence of
nonradial motions.

The best estimates of the distance rely on sharply defined symmetric
structures such as spheres or prolate ellipsoids with sharp edges.  We
capitalize on a geometric trick: the measurement of $\dot\theta$ along
the minor axis of a prolate ellipsoid is ideal for deriving an
expansion distance since no geometric correction for inclination is
required. E25 is the optimum structure for expansion studies
because of the prominent and sharp edges along its minor axis and its
well-measured expansion velocity.  E105 is somewhat more diffuse than
E25, and MS92's Doppler velocity mapping has shown it to be a tilted
circular ring rather than an prolate ellipsoid; however, it serves as a
useful consistency check.

{\it E25:} For the profile method, the best measured expansion rate
of E25's minor axis in the [O~III] images is
$\dot\theta_{prof} = 3.63\pm0.82$ mas yr$^{-1}$.  Adopting MS92's
expansion velocity of 16.4 km s$^{-1}$ from the kinematic model
described earlier in this section, and a 10\% velocity uncertainty results
in a distance D$_{E25} = 953\pm235$ pc.  Based on a size of $\theta$ =
3\farcs6 for the minor half-axis of E25, the kinematic age is T$_{E25} =
990\pm223$ yr.

The radial fitting method was used on both the [O~III] and the [S~II]
images. Using the IRAF command ``splot'', Gaussians were fitted to measure
the shift of the fiducial features between the two epochs.  For
[S~II], the radial cuts were offset slightly to avoid a bright cosmic
ray $\sim$ 1\farcs5 to the south of the central star.

We measured an angular expansion rate $\dot\theta_{rad} = 3.36\pm0.86$
mas yr$^{-1}$ for the minor axis of E25 using this method.  The result
agrees to within 10\% with the result of the profile method.  The
kinematic model described above implies a distance D$_{E25} = 1030\pm283$
pc.  The corresponding kinematic age T$_{E25} = 1070\pm274$ yr.

The magnification method gives very consistent results. The expansion
of most of the inner core (within the intersection of E25 and E105) is
well characterized by a single magnification factor midway
between 1.0025 and 1.003. We adopt a value of 1.00275 and an error of
25\% ({\it i.e.}, $\dot\theta_{mag} = 3.4\pm0.9$ mas yr$^{-1}$).
Judging from the darkened residuals along E25, there may have been a
slight decrease in surface brightness as well.  The distance derived
from the magnification method results is D$_{E25} = 1021\pm288$ pc.  We
derive a kinematic age T$_{E25} = 1057\pm280$ yr.

Combining the results for the minor axis of E25 with equal weights and
averaging the errors, we derive a best-estimate distance
$\bar{D} = 1001\pm269$ pc and a kinematic age $\bar{T} = 1039\pm259$ yr.
This conservative approach of averaging our uncertainties together is 
warranted because our measurements all rely on the same images and radial
velocities, and are therefore not statistically independent.

Next we consider the measurement of $\dot\theta$ along the major axis of
E25.  The structure at the ends of the major axis is much more poorly
defined than along the minor axis. Accordingly, the various techniques
produce somewhat scattered results.  We measure 4.51, 1.95 and $\sim$5
mas yr$^{-1}$ using the profile, radial, and magnification methods
respectively.  The disagreement in the results is substantial.  However,
the important point is that on average all of the methods fall about a
factor of two short of the expected result for uniform (``Hubble
Law''-like) expansion, 7.7 mas yr$^{-1}$.  In other words, the kinematic
age of the major axis of E25 is $\sim2000$ yr.

Similarly large expansion ages are obtained not just along the major
axis of E25 but throughout the periphery of the core of NGC~6543 by
almost all methods.  Whether the peripheral structures are truly
older than the material along the minor axis of E25, or whether the
outer material is coeval and decelerated is an issue that will be
addressed in \S4.

In principle, the major axis of E25 can be exploited to give a check on
the expansion distance since we are able to measure its expansion rate
$\dot\theta$.  After some consideration we decided this would be a mistake
for several reasons: (1) $\dot\theta$ must be corrected for inclination,
and the angle is uncertain; (2) the structure of E25 along its major axis
is irregular and faint; (3) our data show that the ellipse E25 does not
expand uniformly, as MS92 assumed, and that the major axis deviates the
most from the uniform expansion pattern; and (4) there is some doubt that
the tips of E25 are actually part of the same physical structure as the
minor axis (Balick, Wilson, and Harrington, in progress).

{\it E105:} E105 has almost the same projected size and shape as E25, so
it is tempting to use it to derive an independent distance to NGC~6543.
However, E105 is lumpier and more diffuse, and measuring its size and
expansion rate is commensurably more uncertain.  In addition, its
expansion velocity is not as well known.

All three methods again yield consistent results on the expansion rate
of E105 along its major and minor axes.  Moreover, those results
concur with the expansion rates derived for E25.  That is, the values
of $\dot\theta$ for E105 follow a uniform Hubble-law expansion relation
based on the angular size and $\dot\theta$ that we measured for the minor
axis of E25.  This is comforting.  This means that E105 and the minor axis
of E25 share the same kinematic ages.

Let us now employ the MS92 model wherein E105 is a planar circular ring
which surrounds E25 and shares the same symmetry axis.  Since E105 is a
tilted circular ring, the ring's expansion rate measured along its major
axis, $7.1\pm1.8$ mas yr$^{-1}$, needs no geometric correction for tilt.
(This value is an average of the results of the various methods in
Table~2.)  The measured expansion velocity, is 28 km s$^{-1}$ (MS92), and
the derived distance $D_{E105}$ is $832\pm211$ pc, where the errors in the
expansion velocity (which may be substantial for this feature) have been
ignored.  Satisfyingly, D$_{E105}$ is consistent with $\bar{D}$.

{\it Caps $D-D^\prime$:} The caps are lumpy complexes which lie within
thin regions in the outermost extremities of the nebular core.  The
locations of the caps at the extreme edges of the core suggest that
the pair of them lie in or near the plane of the sky.  This means that
their Doppler shifts should be, in principle, unmeasurable, and no
distance can be estimated from them.  On the other hand, this same
geometry guarantees a straightforward computation of the kinematic age
once the angular expansion rate of the caps has been measured.

If NGC~6543 expands uniformly at the rate determined from the minor
axis of E25 then we would expect the caps to exhibit angular expansion
rates $\ga$ 10 mas yr$^{-1}$.

Using the profile method, we detected the signature of angular
expansion from the polar caps in both [O~III] and [S~II].  In the case
of [O~III], the expansion rate  of the polar caps averages to
6.70$\pm$1.37 mas yr$^{-1}$.  In the case of [S~II], the average expansion
of the polar caps is $6.94\pm1.42$ mas yr$^{-1}$. The discrepancy from
uniform expansion lies well outside of measurement errors.

These results are verified by the radial method, which gives expansion
rates of $6.56\pm1.3$ mas yr$^{-1}$ and $5.86\pm1.2$ mas yr$^{-1}$ for
[O~III] and [S~II], respectively.

The residuals of the caps are minimized for values of $1.0015\la M
\la1.0020$ in both the [S~II] and [O~III] images.  Our value of $M =
1.00175$ corresponds to $\dot\theta$ of $6.4\pm1.6$ mas yr$^{-1}$, which
agrees very well with the results from the other two methods.

When all three methods are averaged together, the caps have an expansion
rate of $6.46\pm1.4$ mas yr$^{-1}$, which is more than a third smaller
than the  expected uniform expansion rate.  They are characterized by a
kinematic age of $1628\pm375$ yr.  The caps, like the major axis of E25
have significantly larger kinematic ages than the inner regions of E25.

{\it Condensations $F-F^\prime$, \& Jets $J-J^\prime$:} If the
condensations follow the expansion rate of the minor axis of E25 then
we expect to measure a large angular expansion rate, $\sim12$ mas
yr$^{-1}$. Yet their proper motions are barely discernible (Table 2 and
Figure 5).  This is even more peculiar in light of their relatively
large Doppler shifts, $\sim \pm$25 km s$^{-1}$ (MS92) which, after
(presumably) large correction for inclination suggest true space
velocities $\sim42$ km s$^{-1}$ relative to the nucleus.  It is
therefore quite curious that their proper motions are so small.

Unfortunately the roll angle of HST excluded the interesting polar
jets J-J$^\prime$ from the 1997 images. Future HST observations of
NGC~6543 using the 1994 roll angle should allow measurement of their
proper motion.

{\it Decreasing Expansion Rates.} All of the methods seem to agree
that the apparent expansion rates do not increase linearly with
increasing radius, {\it i.e.} the core of NGC~6543 does not expand
uniformly.  Either the outer nebular gas is decelerating, or it was
ejected considerably earlier than was the gas within $\sim 6\arcsec$
of the nucleus.  We shall return to these points later.

\subsection{Discussion of Distances and Methodologies}

Based primarily on E25 we have derived a distance to NGC~6543 of
1001$\pm$269 pc.  This result is consistent with the previous distance
determinations of 1100, 1170, and 980 pc by Cudworth (1974), Castor
{\it et al.}\ (1981), and Cahn {\it et al.}\ (1992), respectively, but
is discrepant from the 640 pc distance computed by Daub (1982) by about
1.4$\sigma$.

The various methods that we used to measure angular
expansion rates all appear to be feasible.  (The glaring exception is
$\dot\theta_{rad}$ measured for the E25 major axis in
[O~III] line.) Indeed, based on the experience that we gained, it seems
that proper motions of 0.25 pixels, or about 12 mas with the PC camera,
can be measured with errors of order 20\% when the signal is strong.
This applies not only to images obtained under ideal conditions (same
guide stars, centering, spacecraft orientation) but under situations in
which substantial regridding of the data and corrections for geometric
distortions and cosmic rays are required.

According to the papers by Hajian, Terzian, \& Bignell (1993, 1995)
and Hajian \& Terzian (1996), the profile method might even achieve
accuracies of 2 mas under better conditions (sharper features, better
control of the observations, ideal exposure durations, etc.)  than
those in this study.  However, the ultimate technique would exploit
the two dimensions of the image data, perhaps based on the optimum
reduction of the residuals along various filamentary features or the
nebula as a whole.  Of course, at some point expansion parallax
studies will be limited by systematic errors in the WFPC2 detectors,
but apparently this limit has not been reached for motions as small a
few mas.

However, the most severe limitation of any PN expansion distance method
remains a
combination of assumptions about the three-dimensional geometry of the
target and the measurement of expansion velocities, especially with
spatial resolutions obtainable from the ground in complex targets such
as NGC~6543.

\section{Probing the Evolution of NGC~6543}

MS92 built a heuristic (empirical) model for the evolution of NGC~6543
from the best (ground-based) images and their detailed long-slit
spectroscopy available in 1991.  They assumed that each feature in the
core of the nebula was axisymmetric and uniformly expanding.  From the
data and this assumption they could derive the dimensions and inclination
of each feature.

E25 was modeled as a closed, prolate ellipsoid, and E105 is a
circular ring surrounding E25 in its equatorial plane.  In their
model a second, or ``outer'' thick ellipsoid surrounds the nebular
core.  Its inner edges are approximately delineated by the caps
$D$ and $D^\prime$ on its major axis and the projected tips of E105
along its minor axis.  MS92 found that both E25 and the outer
ellipsoid expand uniformly along nearly the same symmetry axis and
inclination.

The newer and more detailed HST images require an entirely new geometric
interpretation. Our heuristic model for the geometry of NGC~6543 is shown
in Figure 6. We retain the idea of E25 as a prolate ellipsoid, albeit
with numerous small bulges and irregularities near its tips.  We are
confident that E25 is a prolate ellipsoid for two reasons, (1) MS92's
spectroscopic data showed with high certainty that it is a closed
ellipsoid, and (2) the HST images are highly suggestive of a prolate
ellipsoidal structure.  The outer thick ellipsoid is replaced with a
figure-eight outline of two roughly-spherical bubbles with truncated
outer regions where the caps are seen.  E105 is situated at the locus
where the bubbles are conjoined at the waist, much like a figure
eight with a fat waist, or ``fat eight'' ({\tt8}).  E25 forms one
coherent structure, and E105 and the fat eight together form an
ensemble which we denote the ``Conjoined Bubbles''.

Both E25 and the Conjoined Bubbles have sharp edges which are nearly
unresolved.  If they are $\bar{T} \approx1000$ yr old or older, and if
these sharp edges (which are strong nebular pressure gradients) were
free to relax (expand) at their internal sound speed of 10 km s$^{-1}$
then they would be $\ga$3 x 10$^{16}$ cm, or $\ga$2\arcsec, in
diameter.

However, almost all of the features in the core of NGC~6543 have edges
which are $\la0\farcs2$ thick, suggesting that they are constrained by
some pressure, perhaps thermal or ram pressure. What's more, if
the edges of the expanding features have been accreting material with low
specific momentum then they will have been decelerated.  The further they
have traveled the more they will have slowed, nicely in accord
with the expansion patterns discussed in \S3.2. In any event, the growth
patterns of the core of NGC~6543 provide a very interesting and useful
glimpse into the physics of its formation and evolution, as we discuss
next.

{\it E25.} The standard concept of E25 as a wind heated, thermally
expanding bubble moving forward supersonically ($\sim$ Mach 1.5)
is compatible, on the whole, with its morphology, measured Doppler shifts,
and pattern of angular expansion.  The gas which the bubble has displaced
and swept up is now compressed into the bright rim which outlines it.  The
leading shell seems to have developed thin-shell instabilities, or lumps,
of small amplitude during its evolution.  This happens because the outward
flow of the thermally-driven expanding bubble surface encounters dense
regions upstream and bends around them, forming surface lumps.  (However,
the scale of the lumps is too small to allow studies of their
individual expansion patterns.)

The 1994 and 1997 observations show that although E25 is likely to be
a prolate ellipsoid, it is not expanding uniformly.  Indeed, its
expansion rate is smallest at its outermost tips.  The expansion of
the Conjoined Bubbles is more difficult to ascertain; however, it's
angular expansion rate is roughly that of E25.  Therefore, the
Conjoined Bubbles are either older or more decelerated than E25, or
both.

Lumpy ellipsoids like E25 are common in elliptical planetary nebulae
such as NGC~6543.  Close counterparts are seen in HST images of
NGC~5882, 6826, 6884, and especially NGC~7009 (Balick {\it et al.}\
1998).  In the last case the bubble has essentially the same
morphology as does E25.  Emanating from the tips of the bubble in
NGC~7009 are highly collimated gas jets with bright ansae at their
termini.  The symmetry axis of NGC~7009 appears to be oriented close
to the plane of the sky.  However, if viewed at an inclination of
$\sim30^\circ$, the bubble, jets, and ansae would resemble E25 and the
jets J-J$^\prime$ in NGC~6543.

The elongated shape of E25 and the bubble of NGC~7009 are generally
believed to be the consequence of a high-pressure waist of confining
material that forces the thermally-expanding bubble to grow most
rapidly along the direction orthogonal to the disk ({\it e.g.}, Icke
{\it et al.}\ 1992).  Thick disks or tori produce prolate ellipsoids
because the expansion in a large equatorial region is hindered.  Thin
disks on the other hand, produce pairs of figure-eight lobes as in
the homunculus of $\eta$ Carinae (Morse {\it et al.}\ 1998) and the
Conjoined Bubbles of NGC~6543, because the restricting material is
confined to a narrow, dense region, while allowing free expansion in the
rest of the nebula.

The concept of E25 as a thermally expanding bubble fails two tests.
Firstly, the torus that putatively constrains the equatorial growth of E25
should be situated astride the minor axis of E25 and probably interior to
E105.  Such a torus is readily discerned in NGC~7009.  However, no such
torus is apparent in NGC~6543, perhaps owing to the confusion of brighter
features nearby.

Secondly, the prolateness develops because the wind-driven bubbles
expand fastest in the polar directions where the upstream confining pressures
are smallest.  Therefore the angular expansion rate of E25 is
predicted to be largest along its major axis.  Rather, just the
reverse is seen.  The kinematic age of the tips of E25 is
roughly twice as large than the kinematic age of its waist.  If the
observed trends continue, the E25 ellipsoid will evolve to become
increasingly less prolate in time.

To rescue the concept of E25 as a thermally expanding bubble one might
argue that although the tips of the bubble originally expanded much
faster than its waist, the tips have now displaced and accreted
considerable amounts of gas with lower specific momentum and have been
decelerated.  Alternately the gas behind the tips may have cooled
adiabatically, so the thermal pressure driving the expansion has
abated.

{\it Conjoined Bubbles.} Almost certainly the Conjoined Bubbles
represent the projected edges of a pair of bubble-like shells
conjoined along their waist. Along its waist a ruffled and slightly
irregular series of ragged low-ionization, elongated knots (E105)
which seem to point away from the nucleus.  These knots may be the
result of an instability formed as the edges of the expanding bubbles
pinch and compress the trapped gas in their waist, like the blades of
a pair of scissors.  Alternately they are ablation flows from dense
neutral knots subjected to ionizing radiation (Mellema {\it et al.}\
1998; Redman
\& Dyson 1999).

Although E25 has counterparts in many other planetaries, the Conjoined
Bubbles do not.  The closest analogues may be NGC~7027 and some
bipolars such as NGC~650-1, 2371-2, and 6309. Interestingly, the Conjoined
Bubbles also mimics $\eta$ Carinae, except that the waist of the latter
object looks like a large thin disk with radial streaks, and its
figure eight is more pinched.  If $\eta$ Carinae follows the pattern
of decreasing expansion with radius that presently characterizes the
Conjoined Bubbles, then it is not difficult to imagine that its overall
morphology will evolve over time to resemble the Conjoined Bubbles!
Dwarkadas \& Balick (1998) showed how this might happen if the
hydrodynamic evolution of the system is dominated by thermal expansion and
adiabatic cooling.

{\it Historical Evolution.}  One key fact is that the waist of the
Conjoined Bubbles, E105, and E25 have the same kinematic age within our
error bars. Hence E25 and the Conjoined Bubbles formed at
roughly the same time.  Based on the increasing kinematic age with radial
distance from the nucleus, it appears that E25 is overtaking some of the
more distant nebular regions such as the tips of E25 and the caps
D-D$^\prime$.

We propose a fairly simple framework for understanding the evolution of
the core of NGC~6543.  About 1000 yr ago, the star ejected a pulse of
material which now forms the conjoined bubbles.  However, the absence of
limb brightening along the edges of the conjoined bubbles (except at $D -
D^\prime$) suggests that the ejection event filled the interior with
low-density gas.  For example, this might have happened if the gas formed
a shock which left some of the outflowing gas in a hot state.

Starting shortly after the pulse of mass ejection the nucleus has been
blowing a high-speed wind, perhaps much like the wind observed today:
$\dot{M}_{wind}$ $\sim 10^{-7.4}$ M$_{\odot}$ yr$^{-1}$ and $v_{wind}
\sim 1900$ km s$^{-1}$ (Perinotto, Cerruti-Sola \& Lamers 1989).  This
wind has swept into the interior of the expanding Conjoined Bubbles.  It
has displaced and plowed the upstream gas and has formed the feature E25:
therefore, E25 defines the ``realm'' of the wind-affected portion of the
core of the nebula.

E25 is now seen as a prolate balloon pushing into the interior of the
Conjoined Bubbles, fastest near the equator, and decelerating towards the
poles.  As we noted before, the plowing of slower upstream gas and
adiabatic cooling inside the tips may have slowed the expansion of E25
along its major axis.

The upstream portions of the Conjoined Bubbles have yet to know of the
existence of E25.  Thus the morphology of the Conjoined Bubbles is a
result of events that shaped it initially.  This feature ``remembers'' its
origins.  Its pinched waist suggests that a thin, dense disk shaped it into
a bipolar when it was formed.  E105 is the remnant of this disk.

This fanciful scenario fails to account for the symmetrically placed
caps, condensations, and jets of NGC~6543.  We can only conjecture
that the caps are dense, neutral material left behind by an even
earlier ejection of mass.  Their low specific momentum and neutral,
dense gas form impediments to the polar growth of the Conjoined Bubbles
and, at the same time, cause dense ionization fronts to form.  The origins
of the condensations and jets are even less obvious.  We shall return
to these issues in a later paper in which the detailed ionization
structure of the nebula is discussed.

\section{Summary}

The excellent agreement of each of the three methods presented
confirms that we have, for the first time, measured the optical
expansion parallax of a PN, and that our error estimates are adequate.
We have shown that accurate measurements of the expansion of nebular
features are readily obtainable with the WFPC2 camera with a reasonable
time baseline. We have further shown that the E25 major axis, the polar
caps, the polar condensations, and the Conjoined Bubbles are all expanding
significantly slower than expected of a Hubble-like expansion. This result
is confirmed in both the [O~III] and [S~II] images, and indicates a region
in which ejected gas is either decelerating, or left behind from an
earlier epoch, or both.

Combining the proper motion of bright fiducials of the nebula with
corresponding spectroscopically obtained radial expansion velocities
(MS92) has allowed us to determine the distance to NGC~6543 of
1001$\pm$269 pc.  Our deduced kinematic age of the brightest inner
parts of the core of the nebula, based on three measurements along the
minor axis of the tilted prolate ellipsoid E25, is 1039$\pm$259 yr.
The outer regions, which resemble a pair of conjoined bubbles with a
large waist, have longer kinematic ages.

A heuristic and highly conceptual explanation of the evolution of these
features is as follows.  A major outburst occurred a thousand years before
today's nebula formed.  Even earlier outbursts produced the huge halo that
surrounds NGC~6543, the polar caps, and other features beyond about a
radius of 20$^{''}$ from the nucleus.  These debris form the environment
for the latest ejection.

The major outburst formed a dense equatorial disk, the remnants of which are
E105.  In addition, the ejected gas was shaped by this disk into a pair of
expanding bubbles which join at E105.  The ``Conjoined Bubbles'' were
created in such a way that their interiors are uniform and probably in a
hot state.  These bubbles are expanding and adiabatically cooling.

Stellar winds began to blow at high speeds shortly after the latest massive
outburst.  These have created an overpressured hot bubble which expends
into the interior of the conjoined bubbles.  The bubble attained a prolate
geometry by rapid expansion along its polar direction.  The subsequent
evolution of the bubble has formed E25.

Like many other interior bubbles in planetary nebulae, E25 is forming lumps
on its surface, perhaps the result of the buckling of its thin expanding
shell.  The expansion of the tips of the prolate bubble has displaced and
accreted material of lower specific momentum, so the tips have
decelerated much more than its waist.  E25 is therefore becoming less
prolate as it expands.

In the future E25 may overtake the conjoined bubbles.  When this occurs
Rayleigh-Taylor instabilities will set in, shredding the remnants of the
nebula, and forming a clumpy halo perhaps much like the larger one that
presently surrounds NGC~6543 ({\it e.g.,} Balick {\it et~al.} 1993).

The ongoing GO 7501 program has selected 30 PNe for WFPC2 observations
for its ongoing multi-epoch expansion parallax program. First epoch
observations are nearly complete.  Subsequent observations will allow
accurate expansion parallax distance measurements of these PNe.  Third
epoch observations of NGC~6543 are planned for late in the year 2000
during which the observing configuration of the 1994 observations by
Harrington \& Borkowski (GO 5403) will be replicated in the bright,
complementary nebular lines of [O~III] and [N~II] as closely as
possible.  This observation should provide excellent data spanning
six years under highly controlled conditions, allowing even more
accurate measurements of the expansion patterns and the distance to
NGC~6543.

\acknowledgments

We are extremely grateful to Virginia Player for her diligent efforts
in making astrometric measurements for our analyses. We are most grateful
to the Space Telescope Science Institute for providing archived data
used in this study.  Support for this work was provided by NASA
through grant number GO 7501 from the Space Telescope Science
Institute, which is operated by AURA, Inc., under NASA contract NAS
5-26555.

\begin{deluxetable}{rlrrcccl}

\footnotesize
\tablecaption{WFPC2 observations of NGC~6543. \label{tbl-1}}
\tablecolumns{8}
\tablewidth{0pc}
\tablehead{
\colhead{Filter} & \colhead{Exp Time} &
\colhead{RA} &
\colhead{Dec} &
\colhead{HST P.A.} &
\colhead{Date}    & \colhead{Prop ID}   & \colhead{PI} }
\startdata

[O~III] (F502N) & 2x200s, 2x600s & 17$^{h}$ 58$^{m}$ 28$\fs08$ &
66$^{\circ}$37\arcmin $59\farcs26$ & 274.105$^{\circ}$
& 9-18-94 & 5403 & Harrington \nl
[O~III] (F502N) & 200s & $17^{h}$ 58$^{m}$ 28$\fs91$ &
66$^{\circ}$38\arcmin
14$\farcs82$ & 305.434$^{\circ}$
& 8-17-97 & 6943 & Casertano \nl

[S~II] (F673N) & 2x400s & 17$^{h}$ 58$^{m}$ 28$\fs08$ &
66$^{\circ}$37\arcmin 59$\farcs26$
& 274.105$^{\circ}$
& 9-18-94 & 5403 & Harrington \nl

[S~II] (F673N)  & 400s & 17$^{h}$ 58$^{m}$ 28$\fs91$ &
66$^{\circ}$38\arcmin
14$\farcs82$ &
305.434$^{\circ}$ & 8-17-97 & 6943 & Casertano \nl

H$\alpha$ (F656N) & 200s & 17$^{h}$ 58$^{m}$ 28$\fs08$ &
66$^{\circ}$37\arcmin 59$\farcs26$
& 274.105$^{\circ}$
& 9-18-94 & 5403 & Harrington \nl

H$\alpha$ (F656N)  & 200s & 17$^{h}$ 58$^{m}$ 28$\fs91$ &
66$^{\circ}$38\arcmin
14$\farcs82$ &
305.434$^{\circ}$ & 8-17-97 & 6943 & Casertano \nl

\enddata

\end{deluxetable}

\begin{deluxetable}{lclll}

\footnotesize
\tablecaption{Angular Expansion Rates of NGC~6543.  All
values of $\dot\theta$ are in units of mas yr$^{-1}$. \label{tbl-2}}
\tablecolumns{5}
\tablewidth{0pc}
\tablehead{
\colhead{Feature} & \colhead{Angular size (diameter)} &
\colhead{Method} & \colhead{$\dot\theta$-[O~III]} &
\colhead{$\dot\theta$-[S~II]}}
\startdata

E25 minor & 7\farcs2 (in [O~III]) & Profile & 3.63$\pm$0.82 &  \nl
 &  & Rad. Fit & 3.36$\pm$0.86 &   \nl
 &  & Mag. & 3.4$\pm$0.9 &  \nl
 & & & & \nl
E25 major & 16\farcs1 (in [O~III]) & Profile & 4.51$\pm$0.85 &  \nl
 & & Rad. Fit & 1.95$\pm$1.01 &  \nl
 & & Mag. & 4.8$\pm$1.2 &  \nl
 & & Linear (Hubble) Expectation & 7.7   &   \nl
 & & & & \nl
E105 minor & 9\farcs0 (in [S~II]) & Profile & 5.8$\pm$1.5 &   \nl
 & & Rad. Fit & 3.98$\pm$1.17 & 3.28$\pm$0.7  \nl
 & & Mag. & 4.6$\pm$1.2 & 4.6$\pm$1.2  \nl
 & & Linear (Hubble) Expectation &  4.3  &   \nl
 & & & & \nl
E105 major & 14\farcs6 (in [S~II]) & Profile & 7.2$\pm$1.5 &   \nl
 & & Rad. Fit & 4.68$\pm$2.34 & 8.59$\pm$1.7 \nl
 & & Mag. & 7.5$\pm$1.9 & 7.5$\pm$1.9  \nl
 & & Linear (Hubble) Expectation & 7.0  &   \nl
 & & & & \nl
Polar Caps (DD$^{\prime}$)& 21\farcs2 (in [S~II]) & Profile &
6.70$\pm$1.37 & 6.94$\pm$1.42 \nl
 & & Rad. Fit & 6.56$\pm$1.3 & 5.86 $\pm$1.2  \nl
 & & Mag. & 6.4$\pm$1.6 & 6.4$\pm$1.6 \nl
 & & Linear (Hubble) Expectation &  10.2  &   \nl
 & & & & \nl
Condensations (FF$^{\prime}$) & 24\farcs7 (in [S~II]) & Rad. Fit &  &
2.73$\pm$1.64 \nl
 & & Mag. &  & 4.2$\pm$1.1 \nl
 & & Linear (Hubble) Expectation &   11.9  &   \nl

\enddata

\end{deluxetable}

\vfill
\eject
\def\par{\endgraf \hangindent 15pt
\parindent=0pt}
\hsize 6.5truein
\vskip20pt
\noindent{\bf Bibliography}

Balick, B., Gonzalez. G, Frank, A. \& Jacoby, G.J. 1993, ApJ, 392, 582

Cahn, J.H., Kaler, J.B. \& Stanghellini, L. 1992, A\&A Supp.,
94, 399

Castor, J.I., Lutz J.H., \& Seaton, M.J. 1981, MNRAS, 244, 521

Ciardullo, R., Jacoby, G.H., \& Ford, H.C. 1988, PASP, 100, 1218

Cudworth, K.M. 1974, AJ, 79, 1384

Daub, C.T. 1982, ApJ, 260, 612

Dwarkadas, V., \& Balick, B. 1998, AJ, 116, 829

Gomez, Y., Rodriguez, L.F., \& Moran, J.M. 1993, ApJ, 416. 620

Hajian, A.R., Terzian, Y., \& Bignell, C. 1993, AJ, 106, 1965

Hajian, A.R., Terzian, Y., \& Bignell, C. 1995, AJ, 109, 2600

Hajian, A.R., \& Terzian, Y. 1996, PASP, 108, 258

Harrington, J.P. \& Borkowski, K.J. 1994, Bull.\ Amer.\ Astron.\ Soc.,
26, 1469

Haryadi, C., \& Seaquist, E.R. 1998, AJ, 115, 2466

Heap, S.R., \& Augensen, H.J. 1987, ApJ, 313, 268

Icke, V., Balick, B., Frank, A. 1992, A\&A, 253, 224

Masson, C.R. 1986, ApJ Lett, 302, L27

Masson, C.R. 1989a, ApJ, 336, 294

Masson, C.R. 1989b, ApJ, 346, 243

Mellema, G., Raga, A.C., Canto,J., Lundqvist,P., Balick,B., Steffen, W.,
Noriega-Crespo, A. 1998, A\&A, 331, 335

Miranda, L.F., \& Solf, J. 1992, A\&A, 260, 397 (MS92)

Morse, J.A., Davidson, K., Bally, J., Ebbets, D., Balick, B., \& Frank,
A. 1998, AJ, 116, 2443

Perinotto, M., Cerruti-Sola, M., \& Lamers, H.J.G.L.M. 1989, ApJ, 337,
382

Redman, M.P., \& Dyson, J.E. 1999, MNRAS, 302, 17

Shklovsky, I.S. 1956, AZh, 33, 22

Terzian, Y. 1993, Planetary Nebulae, IAU Symposium \# 155, eds.\
R.\ Weinberger \& A.\ Acker (Dordrecht, Kluwer), p.\ 109

Terzian, Y. 1997, IAU Symposium \# 180, eds.\ H.J.\ Habing \&
H.J.G.L.M.\ Lamers (Kluwer)

\newpage

{\bf Figure Captions:}

{\bf Figure 1:} Images of NGC~6543.  Left: A linear representation of
the [O~III] surface brightness.  Various features are identified (MS92
terminology shown in italics).  White dots show the locations of the
radial fits of angular expansion rates of Table 2.  Center: A negative
linear representation of the [O~III] surface brightness superimposed
on a logarithmic display of the faint nebular background in the same
HST image.  Right: Like the center panel, but for the [N~II] line.
The [S~II] image is noisier but otherwise similar in appearance to
this panel.

{\bf Figure 2:} A linear representation of the [O~III] surface
brightness of NGC~6543 showing the location of the profiles analyzed
in \S 3.2.  Values of the angular expansion rate in mas yr$^{-1}$ are
shown corresponding to the locations (black dots) where the profile
cuts intersect bright regions of the nebula.

{\bf Figure 3:} Same as for Figure 2, except for the [S~II] surface
brightness of NGC~6543.  The values listed correspond to the measured
angular expansion rates in mas yr$^{-1}$ where the profiles intersect
the bright polar caps.

{\bf Figure 4:} A montage of F502N ([O~III]) residual images after the
1997 images were subtracted from magnified and aligned 1994 images.
The magnification factors that were applied to the 1994 images are
shown in the upper right corners of each panel.

{\bf Figure 5:} Like Figure 2 except for the F673N ([S~II]) images.

{\bf Figure 6:} Schematic of a conceptual geometric model of NGC~6543.
A plan view of the model is shown on the right side.  How it might
appear if tilted at about 30$^\circ$ is shown in projection on a
combined [N~II] and [O~III] image.

\end{document}